%% file: main.tex
\documentclass[10pt]{wlscirep}
\usepackage[utf8]{inputenc}
\usepackage[T1]{fontenc}
\usepackage{adjustbox}
\usepackage{subfigure}

\title{Auditing Contextual Bias in Human Ball-Strike Calls Using KBO's Automated Umpiring Transition}

\author[1]{Kichang Lee}
\author[1]{JeongGil Ko}
\affil[1]{Yonsei University, School of Integrated Technology, Seoul, 03789, Republic of Korea}
\keywords{Sports analytics, Baseball, Automatic Ball-Strike System}
\begin{abstract}
\input{sec/0_abstract}
\end{abstract}

\begin{document}

\flushbottom
\maketitle
% \thispagestyle{empty}

\input{sec/1_intro}

\input{sec/2_method}

\input{sec/3_analysis1}

\input{sec/4_analysis2}

\input{sec/5_analysis3}

\input{sec/6_analysis4}

\input{sec/7_analysis5}

\input{sec/8_relwork}

\input{sec/9_discussion}

\input{sec/10_conclusion}

\bibliography{reference}
\end{document}

%% file: sec/0_abstract.tex
This paper uses the Korean Baseball Organization's adoption of the Automated Ball-Strike (ABS) system to audit long-standing claims about contextual bias in human ball-strike calls. Using pitch-level KBO data from 2021 through the available portion of the 2026 season, we model called-strike probability for taken pitches near the strike-zone boundary, with 2022-2023 as the primary human-umpire baseline and ABS seasons (2024 and onward) as a diagnostic benchmark. The strongest evidence concerns count pressure. Relative to 0--0 counts, human umpires called substantially fewer strikes in two-strike counts and more strikes in hitter-ahead three-ball counts. Specifically, in the main 0.25-ft boundary band, 0--2 was associated with a -17.17 percentage-point effect and 3--0 with a +6.61 percentage-point effect. Under ABS, the corresponding effects were close to zero and did not survive false-discovery-rate correction. Game progression shows a smaller but coherent pattern as human calls were less strike-prone in early innings and more strike-prone in innings 7--9+, especially in late-close situations, while complete ABS seasons were essentially flat. Other suspected biases are weaker or more localized. Salary-based reputation proxies provide suggestive but proxy-sensitive evidence, and catcher identity shows human-period residual heterogeneity that disappears under ABS. Home-context evidence is mostly null at the umpire level, with one FDR-significant human-period exception and an exploratory umpire-team gap best treated as an audit lead. Overall, the results do not show that human umpires were biased everywhere. Instead, they map where the human strike zone was most context-sensitive, where evidence was weaker, and where common suspicions received little support.

%% file: sec/1_intro.tex
\section*{Introduction}
\label{sec:intro}

Few judgments in baseball are more rule-bound than the strike zone, yet few have generated more suspicion that context matters. Fans, players, coaches, and analysts have long debated whether umpires call the same pitch differently depending on the game context, for example, count, inning, score, batter, pitcher, or catcher. A borderline pitch in a 3--0 count may feel different from the same pitch in an 0--2 count. A respected batter may be perceived as receiving more benefit of the doubt. A skilled catcher may appear to ``steal'' strikes near the edge. A late-close game may seem to produce a different zone from an early low-leverage inning. These claims are familiar, but familiarity is not evidence.

The difficulty is that contextual umpiring claims are hard to test from human-call data alone. Contexts are not randomly assigned. Pitchers throw different pitches in different counts, batters take different pitches in different situations, catchers are paired with specific pitchers, and stronger players appear in different strategic contexts. A raw difference in called-strike rate can therefore reflect pitch selection, pitch location, pitch type, player quality, or game state rather than a shift in umpire judgment. The empirical problem is not simply to ask whether called-strike rates differ across contexts, but to ask whether comparable borderline pitches are treated differently after pitch geometry and baseball covariates are controlled.

The Korean Baseball Organization's adoption of the Automated Ball-Strike (ABS) system in regular-season play beginning in 2024 creates a rare opportunity to revisit these long-standing claims~\cite{ynaExpandsStrike,dongaKoreanBaseball}. Before ABS, analysts could estimate where human umpires tended to call strikes and balls, but contextual patterns were difficult to separate from pitch selection, player matchups, catcher assignment, and game situation. After ABS, the same league provides an automated benchmark in which ball-strike decisions are no longer made by the home-plate umpire. This transition allows us to ask which suspected contextual patterns were specific to the human-umpire era and which remain even when the call is automated.
We interpret this comparison diagnostically rather than causally because the ABS transition was not a randomized experiment. Patterns that attenuate under ABS are treated as evidence consistent with human context-sensitive judgment, whereas patterns that persist under ABS are interpreted more cautiously as possible pitch-selection imbalance, strategic change, measurement error, or model misspecification.

This paper asks which long-standing suspicions about the human strike zone survive such an audit. Rather than evaluating ABS as a product or asking only whether robot umpires are more accurate, we use the KBO transition to examine whether the practical human strike zone moved with context before automation. We audit five families of claims. First, we test count pressure: whether umpires become more strike-prone in hitter-ahead counts and less strike-prone when a called strike could end the plate appearance. Second, we test player-status effects, using salary as an observable but imperfect proxy for reputation. Third, we examine game progression, including inning, close-game, and late-game situations. Fourth, we estimate catcher- and pitcher-associated residual variation after controlling for pitch location and context. Fifth, we test home context, both as an average home-batter advantage and as more localized umpire--team combinations.

Building on prior work showing that human--ABS discrepancies are concentrated near the strike-zone boundary~\cite{lee2025analyzing}, our design focuses on taken pitches near the rule-zone edge, where contextual judgment should be most visible. We analyze KBO pitch-level data from the human-umpire era and the ABS era, using 2022--2023 as the primary human baseline and 2024--2026 as the automated benchmark. The outcome is whether a taken pitch is called a strike. Across the five audit settings, we compare called-strike probability for similar borderline pitches after controlling for pitch location, zone geometry, pitch characteristics, season, and relevant player or game context.

The results do not support a blanket claim that human umpires were biased in every context. Instead, they produce a hierarchy of evidence. Count pressure is the clearest and strongest signal: before ABS, comparable borderline pitches were more likely to be called strikes in 3--0 counts and less likely to be called strikes in two-strike counts, while the same contrasts were near zero under ABS. Game-progression effects are smaller but structured, with late and close situations showing more strike-prone human calls than earlier contexts. Status-related evidence is more tentative: salary-based proxies suggest some reputation-associated heterogeneity, but the interpretation depends on proxy quality and model specification. Catcher identity is associated with residual called-strike variation under human umpires, and that variation largely disappears under ABS. Home-context evidence is mostly null, suggesting that average home-team advantage is not a major feature of KBO ball-strike calls in this setting.

The contribution of this paper is therefore not to show that automation simply replaced an inaccurate human system. Nor is it to claim that every suspected bias is real. The contribution is an empirical map of contextual structure in human rule enforcement, for instance, where the evidence is strong, where it is modest, and where common suspicions receive little support. In this sense, automated umpiring serves not only as replacement infrastructure but also as audit infrastructure. It provides a benchmark for identifying which parts of the human strike zone were most context-sensitive before the call was automated.

%% file: sec/2_method.tex
\section*{Methodology}
\label{sec:method}

\subsection*{Data Collection}

We analyze pitch-level KBO data collected from publicly accessible Naver Sports game records~\cite{naver}. The local data snapshot contains 1,216,246 pitch rows from 2021 through the available portion of the 2026 season. We use 2022--2023 as the primary human-umpire baseline. The 2021 season is excluded from the primary baseline since the KBO strike-zone environment changed before 2022, but it remains useful for historical sensitivity. ABS was used from 2024 onward. Most ABS benchmark analyses pool 2024--2026 to maximize precision; analyses of game progression use the complete ABS seasons, 2024--2025, as the primary automated benchmark and treat partial 2026 as sensitivity.

\begin{table}[!ht]
\centering
\small
\begin{adjustbox}{width=.8\linewidth,center}
\begin{tabular}{|c|l|r|r|r|r|r|}
\hline
\textbf{Season} & \textbf{Period} & \textbf{Total pitches} & \textbf{Called pitches} & \textbf{Called strikes} & \textbf{Games} & \textbf{Boundary-band called} \\ \hline
2021 & Historical human & 220,345 & 122,660 & 38,559 & 719 & 39,949 \\ \hline
2022 & Primary human & 214,771 & 115,317 & 37,143 & 714 & 37,356 \\ \hline
2023 & Primary human & 215,540 & 117,678 & 38,001 & 708 & 38,617 \\ \hline
2024 & ABS & 222,343 & 120,426 & 38,941 & 718 & 36,705 \\ \hline
2025 & ABS & 217,766 & 118,552 & 39,247 & 720 & 36,082 \\ \hline
2026 & ABS, partial & 125,481 & 69,268 & 22,845 & 408 & 21,267 \\ \hline
\textbf{Total} & -- & \textbf{1,216,246} & \textbf{663,901} & \textbf{214,736} & \textbf{3,987} & \textbf{209,976} \\ \hline
\end{tabular}
\end{adjustbox}
\caption{\textbf{Detailed dataset configuration.} Boundary-band called pitches are taken pitches within 0.25 ft of the nearest rule-zone boundary after strike-zone validation.}
\label{tab:sample_construction}
\end{table}

\subsection*{Called pitches and strike-zone geometry}

The dependent variable is \(\textit{CalledStrike}\), equal to one for a called strike and zero for a called ball. We exclude swings, fouls, balls in play, hit-by-pitch events, pitchouts without a ball-strike adjudication, and other pitches that did not require a taken-pitch call. This restriction is necessary because the paper studies ball-strike adjudication, not swing decisions.

For each called pitch, we construct normalized horizontal and vertical location, rule-zone status, signed distance to the nearest rule-zone boundary, and nearest boundary side. The horizontal rule-zone approximation is \(-0.9\) to \(0.9\) ft. The primary vertical zone uses batter-specific top and bottom strike-zone fields when they pass validation checks; these fields are available for essentially all called pitches in the analysis sample. Because vertical-zone geometry differs across seasons, models include zone-bottom and zone-height controls. The main boundary-band sample contains pitches within 0.25 ft of the nearest rule-zone boundary, where small contextual shifts in the effective zone are most likely to be visible.

\subsection*{Modeling strategy}

Raw strike rates cannot answer the paper's question because called-strike probability changes steeply near the zone boundary. The model therefore first estimates a local strike-zone surface, then asks whether a context variable shifts calls among otherwise comparable taken pitches. This follows prior strike-zone work that treats ball-strike judgment as a probabilistic surface rather than a simple rectangular lookup~\cite{chen2010contour,deshpande2017hierarchical,hunter2018new,flannagan2024psychophysics}.

For pitch \(i\), we estimate logistic models of the form
\[
\Pr(Y_i=1) = \operatorname{logit}^{-1}
\left(\alpha + f_{\mathrm{zone}}(x_i,y_i;b_i,h_i) + \lambda^\top C_i + \gamma^\top X_i + \delta_{s(i)}\right),
\]
where \(Y_i\) is a called strike. The function \(f_{\mathrm{zone}}\) is a parsimonious basis expansion of the local called-strike surface:
\[
f_{\mathrm{zone}}(x_i,y_i;b_i,h_i)
= \boldsymbol{\theta}^{\top}\mathbf{g}_i,
\qquad
\mathbf{g}_i =
\begin{bmatrix}
x_i & y_i & x_i^2 & y_i^2 & x_i y_i & d_i & d_i^2 & r_i & b_i & h_i
\end{bmatrix}^{\top}.
\]
Here \(x_i\) and \(y_i\) are normalized horizontal and vertical locations. \(b_i\) and \(h_i\) are the batter-specific zone bottom and height, and \(d_i=d(x_i,y_i;b_i,h_i)\) is signed distance to the nearest rule-zone boundary. \(r_i=r(x_i,y_i;b_i,h_i)\) indicates whether the pitch is inside the rule-zone approximation. The distance and rule-zone terms are functions of location and zone geometry, not separate physical information. They are included as they encode the rule boundary directly. A model using only smooth functions of \(x_i\) and \(y_i\) must learn the rectangular edge indirectly, whereas signed distance aligns the fit with the nearest boundary, \(r_i\) separates just-inside from just-outside pitches, and the polynomial terms allow smooth curvature and corner asymmetry within that boundary-aligned surface. Thus \(f_{\mathrm{zone}}\) is a compact geometric adjustment designed for borderline pitches, not a claim that each transformed term is a separate causal mechanism.

\(C_i\) is the context being tested in a given analysis, such as count state, reputation proxy, inning phase, catcher identity, or home context. \(X_i\) contains pitch and matchup controls: pitch-type group, standardized pitch speed, batter stance, pitcher hand, and count-pressure controls when count is not itself the treatment of interest. Season fixed effects \(\delta_{s(i)}\) absorb league-wide season differences. We report average marginal effects on called-strike probability in percentage points, with confidence intervals, sample sizes, and false-discovery-rate-adjusted \(q\)-values for figure-level test families. Experiment-specific specifications are introduced with each analysis below.

%% file: sec/3_analysis1.tex
\section*{Results}

\subsection*{Audit 1: Do umpires balance the count?}
\label{sec:analysis_count}

The first audit target is one of baseball's most familiar umpiring claims: umpires do not call the same borderline pitch the same way in every count. Two related suspicions are especially common. In a 3--0 count, umpires may be reluctant to award a walk on a marginal pitch and may therefore call more strikes. In a two-strike count, umpires may be reluctant to end the plate appearance on a marginal pitch and may therefore call fewer strikes. We refer to this family of claims as \textit{count balancing}.

This claim has a direct empirical prediction. The count does not change the rule-book strike zone, but it changes the consequence of the next call. If human umpires balance counts near the boundary, then hitter-ahead three-ball counts should be more strike-prone, while two-strike counts should be less strike-prone after pitch geometry is controlled. Under ABS, the same count states should not produce a comparable ordered pattern. We use 0--0 as the displayed baseline because it is the initial count state and provides a common reference point across all count contrasts.

\begin{figure}[!ht]
\centering
\includegraphics[width=0.8\linewidth]{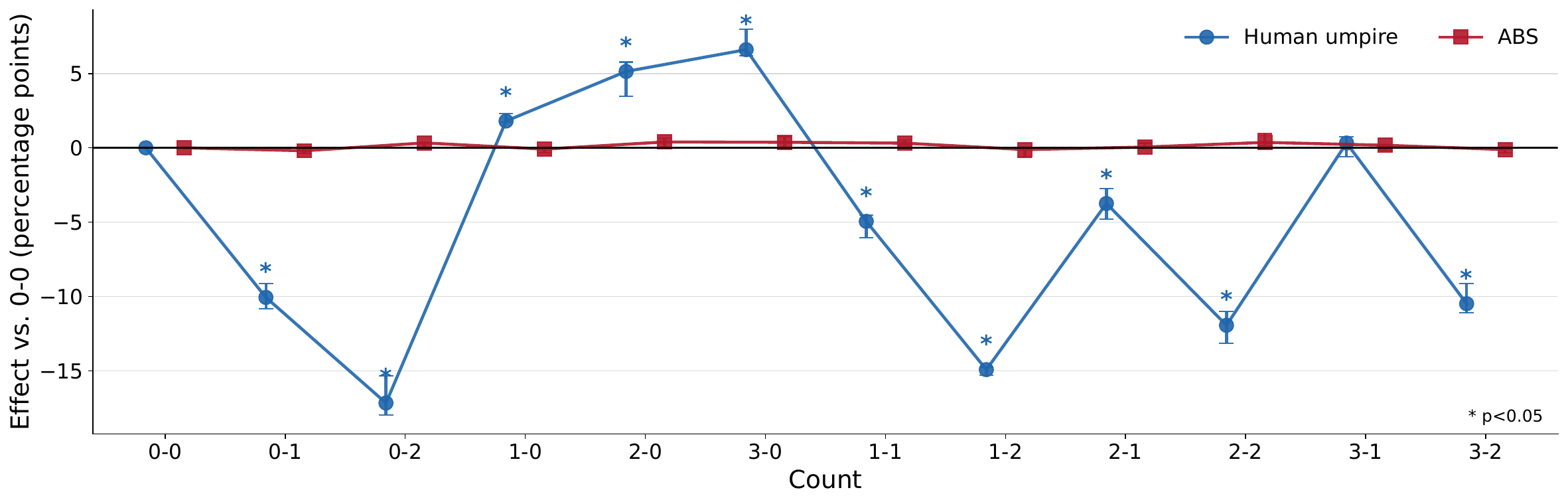}
\caption{\textbf{Count-state effects on borderline called-strike probability.} Effects are average marginal effects in the main 0.25-ft boundary band, with 0--0 as the displayed baseline. Human-umpire estimates use 2022--2023; ABS estimates use the pooled 2024--2026 benchmark. Asterisks indicate false-discovery-rate (FDR) adjusted \(q<0.05\).}
\label{fig:h1_count_effects}
\end{figure}

Figure~\ref{fig:h1_count_effects} shows that the human-period estimates match the count-balancing prediction. Relative to 0--0 counts, 3--0 pitches were 6.61 percentage points more likely to be called strikes. This supports the first part of the claim: marginal pitches became more strike-prone when the alternative was ball four. The opposite pattern appears in two-strike counts. Relative to 0--0, 0--2 pitches were 17.17 percentage points less likely to be called strikes; 1--2 pitches were 14.92 percentage points less likely; 2--2 pitches were 11.94 percentage points less likely; and full-count pitches were 10.49 percentage points less likely. These effects survive FDR correction in the human period.

The pattern is not a generic pro-pitcher or pro-batter tendency. Hitter-ahead counts without two strikes generally move upward, with 2--0 at +5.14 pp and 1--0 at +1.80 pp, while two-strike counts move sharply downward. The exception is 3--1, which is close to zero and not statistically significant. Taken together, the estimates suggest an ordered count-dependent shift: human calls move toward extending the plate appearance rather than allowing a borderline pitch to decide it immediately.

The ABS benchmark provides the diagnostic contrast. In the pooled 2024--2026 ABS period, the corresponding 3--0 and 0--2 effects were only 0.37 pp and 0.33 pp, respectively, and neither survived FDR correction. Other ABS count-state estimates were similarly small. Some ABS point estimates are not exactly zero, which is expected in observational data conditioned on taken pitches, but the ordered human count pattern largely disappears once ball-strike adjudication is automated.

\begin{figure}[!ht]
\centering
\includegraphics[width=0.92\linewidth]{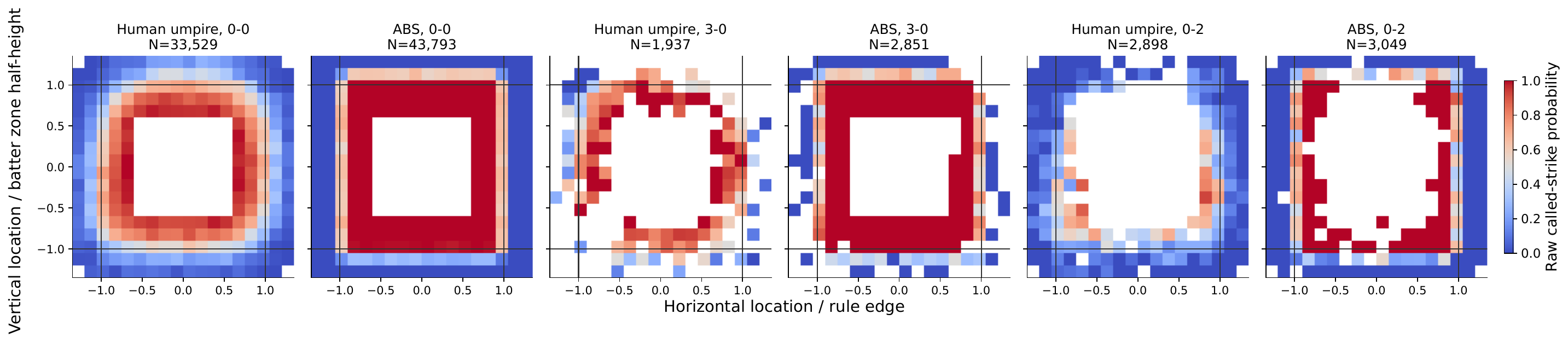}
\caption{\textbf{Count-state surface.} The heatmap summarizes how the called-strike shift varies across all feasible ball-strike counts. The human period shows a diagonal structure: hitter-ahead counts are more strike-prone, whereas two-strike counts are less strike-prone. The ABS benchmark is nearly flat.}
\label{fig:h1_count_heatmap}
\end{figure}

Figure~\ref{fig:h1_count_heatmap} shows the same result as a count-state surface. The human period has a clear count-contingent structure: the upper hitter-ahead region is more strike-prone, while the two-strike region is less strike-prone. The ABS surface is nearly flat, indicating that the same count-state structure is not produced by the automated adjudication benchmark.

The verdict for Audit 1 is strong support for the count-balancing claim. The evidence does not require the claim that umpires consciously manipulated counts. A safer interpretation is that the effective human decision boundary shifted under consequence pressure. Marginal pitches were more likely to become strikes when the alternative was a walk, and less likely to become strikes when the call could end the plate appearance. Among the suspected contextual biases tested in this paper, count balancing is the clearest human-period signature.

%% file: sec/4_analysis2.tex
\subsection*{Audit 2: Do high-status players get a different zone?}
\label{sec:analysis_status}

The second audit target is another familiar baseball claim: established players receive more benefit of the doubt. In ball-strike terms, this claim has a simple directional prediction. High-status batters should receive fewer called strikes on comparable borderline pitches, while high-status pitchers should receive more called strikes. Unlike count pressure, however, this claim immediately raises a measurement problem. Reputation is not directly observed in pitch-level data.

We therefore begin with salary as an observable proxy for player status. Salary is useful because more established, recognizable, and valuable players generally command higher contracts. But salary is not reputation itself. Contract timing, late-career arrangements, voluntary salary reductions, and team-specific decisions can separate a player's recorded season salary from his public baseball status. This limitation is not just theoretical; it becomes visible in the data.

The most revealing case is Shin-soo Choo. In the player-level salary--residual plot, Choo appears as a low-salary batter despite being one of the most recognizable Korean baseball players of his generation.\footnote{For readers unfamiliar with Korean baseball, Shin-soo Choo  is not a typical low-status player. He was a long-time Major League Baseball player and former MLB All-Star; Baseball-Reference credits him with 1,671 MLB hits and 218 MLB home runs. In his final KBO season, however, he signed for the KBO-minimum salary of 30 million won after earning 1.7 billion won the previous year and announced that he would donate the full salary to charity. This makes him a clear case in which recorded season salary understates baseball reputation.} This is precisely the kind of observation that should make a salary-based reputation test difficult. If umpires respond to reputation-like information rather than recorded salary itself, then a famous low-salary veteran will be misclassified by the salary proxy.

We therefore treat the Choo case as a proxy-validity stress test rather than as a nuisance outlier. The question is not whether removing one player can make a coefficient significant. The more substantive question is whether the result changes when the clearest reputation--salary mismatch is accounted for. If the salary proxy is meaningful but noisy, then removing a player whose salary clearly understates reputation should make the salary gradient more aligned with the reputation hypothesis.

\begin{figure}[!ht]
\centering
\includegraphics[width=0.92\linewidth]{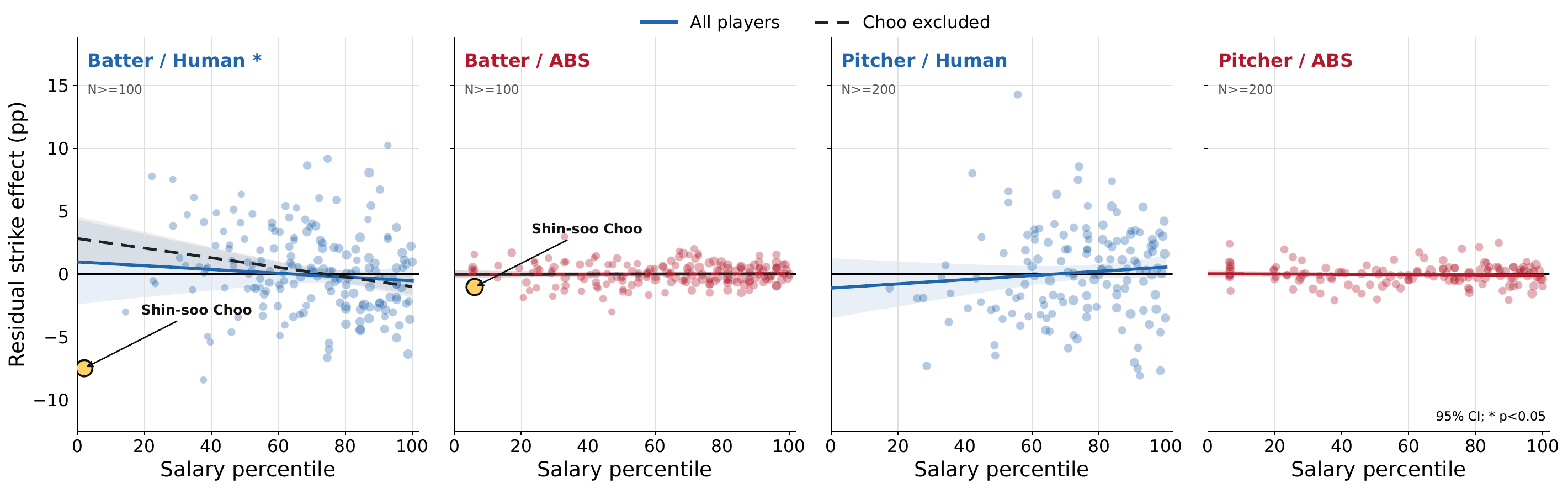}
\caption{\textbf{Player-level residual effects and salary as a reputation proxy.} Each point represents a player-level residual called-strike effect and season salary percentile after minimum-pitch filters. Negative batter slopes indicate that higher-salary batters received fewer called strikes; positive pitcher slopes indicate that higher-salary pitchers received more called strikes.}
\label{fig:h2_salary_correlation}
\end{figure}

Figure~\ref{fig:h2_salary_correlation} shows the identical pattern. Among batters with at least 100 called pitches, the human-period salary slope is negative but not statistically supported when all players are included (-1.50 pp, \(p=0.496\)). After excluding Shin-soo Choo , the slope becomes substantially larger and statistically supported (-3.82 pp, 95\% CI -6.25 to -1.39, \(p=0.002\)). This change is important because it has a baseball explanation. The association becomes visible only after removing a player whose recorded salary is clearly misaligned with his reputation. In baseball terms, the adjusted result matches the reputation claim: higher-salary batters received fewer called strikes after location and context controls.

This does not mean that the Choo-excluded estimate is the only correct estimate, nor does it prove that umpires consciously rewarded famous batters. Instead, the sensitivity shows that the batter-status result depends on construct validity: whether salary is treated as a clean measure of reputation or as a noisy proxy with identifiable mismatches. The all-player model gives a weak batter association; the proxy-stress-tested model reveals a statistically supported association in the predicted direction. Under ABS, the batter slope is approximately zero regardless of whether Choo is included, which makes the human-period sensitivity more informative.

\begin{figure}[!ht]
\centering
\includegraphics[width=0.9\linewidth]{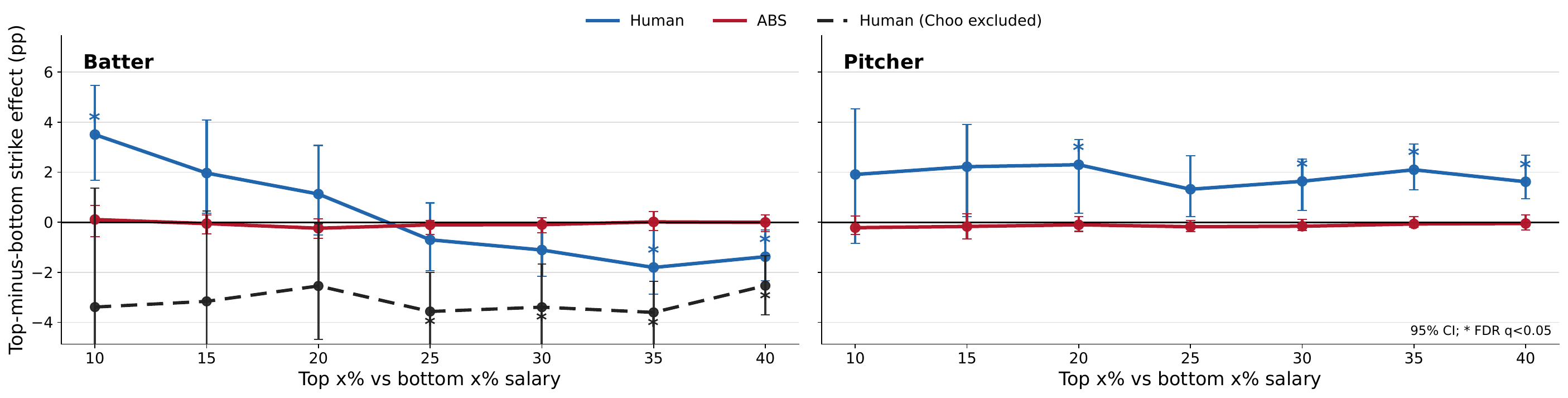}
\caption{\textbf{Salary cutoff sweep by player role.} Points show top-versus-bottom salary contrasts across percentile cutoffs in the main 0.25-ft boundary-band sample. For batters, negative effects favor higher-salary batters; for pitchers, positive effects favor higher-salary pitchers. Asterisks indicate FDR \(q<0.05\).}
\label{fig:h2_salary_cutoff}
\end{figure}

Figure~\ref{fig:h2_salary_cutoff} provides a complementary cutoff-based test. The pitcher results are more stable than the batter results. In the human period, top-versus-bottom pitcher contrasts are positive across several cutoffs, with FDR-supported effects at the 20\%, 30\%, 35\%, and 40\% thresholds and effect sizes between +1.62 and +2.30 percentage points. This direction is consistent with the claim that higher-status pitchers received a more favorable zone. Under ABS, the corresponding pitcher contrasts are close to zero.

The batter cutoff sweep is less stable. The 10\% cutoff is positive, whereas broader 35\% and 40\% cutoffs are negative. This instability is consistent with the proxy problem identified above: mechanical salary groups can mix reputation, contract structure, career stage, and sample composition. The batter cutoff analysis should therefore not be interpreted as a clean monotonic salary effect. It is better read together with the player-level residual plot, where the key issue is whether salary successfully represents reputation.

The verdict for Audit 2 is partial and proxy-sensitive support. The most interesting feature is not a simple salary effect, but the way the result changes when the salary proxy is stress-tested. A clear reputation--salary mismatch weakens the all-player batter association; after that mismatch is removed, the human-period batter slope becomes statistically supported in the predicted direction. Pitcher cutoff contrasts also point in the reputation-favoring direction across several thresholds. These patterns attenuate under ABS. Because reputation is measured indirectly and the batter result depends on a defensible but consequential sensitivity choice, this audit should be treated as suggestive rather than confirmatory.

%% file: sec/5_analysis3.tex
\subsection*{Audit 3: Does the zone change late in games?}
\label{sec:analysis_game}

The third audit target is the familiar ``go-home zone'' claim. In informal baseball language, umpires are sometimes suspected of expanding the strike zone late in games, especially when a game is long, close, or effectively decided. This claim is intuitive but underspecified. A simple go-home story predicts a broad late-game increase in called strikes. A more contextual version predicts that the zone changes only in particular late-game states, such as close games or situations in which the batting team is trailing. We therefore treat Audit 3 as a test of whether the human strike zone drifts over game phase, and whether that drift is consistent with a simple go-home explanation.

The empirical prediction is straightforward. If umpires generally become more strike-prone as games progress, then late innings should show higher called-strike probability than early innings after location and count controls. If the effect is specifically about ending games, then late blowouts should also become more strike-prone. If instead the effect reflects contextual pressure rather than a simple desire to finish the game, the late-game pattern should be concentrated in particular score states and should not necessarily appear in lopsided games. Under ABS, comparable inning-phase patterns should be absent or much smaller.

\begin{figure}[!ht]
\centering
\includegraphics[width=0.8\linewidth]{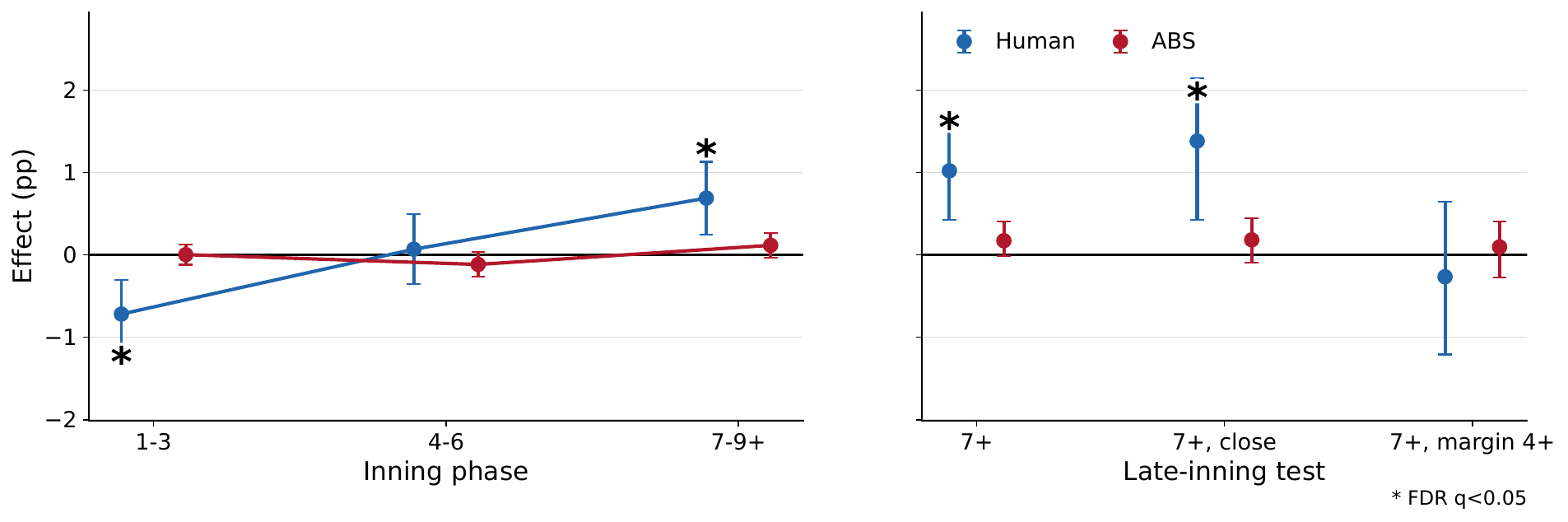}
\caption{\textbf{Inning-phase and late-game context tests.} Left: inning-phase effects, measured as deviations from each period's average called-strike probability after location and count controls. Right: a late-game context family covering all innings 7+, late-close situations, and late blowouts. Human estimates use 2022--2023; ABS estimates use complete 2024--2025 seasons. Asterisks indicate FDR \(q<0.05\).}
\label{fig:h3_late_primary}
\end{figure}

Figure~\ref{fig:h3_late_primary} shows the main game-phase result. In the human period, called-strike probability moves upward as the game progresses. Early innings 1--3 were 0.72 percentage points less strike-prone than the period average (FDR \(q=0.0018\)), middle innings 4--6 were near zero, and late innings 7--9+ were 0.69 percentage points more strike-prone (FDR \(q=0.0033\)). The complete ABS seasons do not show the same drift: innings 1--3 were 0.00 pp, innings 4--6 were -0.12 pp, and innings 7--9+ were +0.12 pp, with no FDR-supported phase effect. Thus, the human period shows a modest but ordered early-to-late shift, while the automated benchmark is essentially flat.

The binary late-game tests sharpen the interpretation. In the human period, all innings 7+ were 1.02 pp more strike-prone (FDR \(q=0.0082\)), and late-close situations were 1.38 pp more strike-prone (FDR \(q=0.0069\)). These results support the idea that late-game context matters. However, the late-large-margin estimate was slightly negative (-0.27 pp) and not statistically supported. This matters because a simple go-home theory would predict more strikes in late blowouts as well. The data instead point to a narrower pattern: human calls became more strike-prone late in games, especially when the game was close, but not simply whenever the game was late.

\begin{figure}[!ht]
\centering
\includegraphics[width=0.85\linewidth]{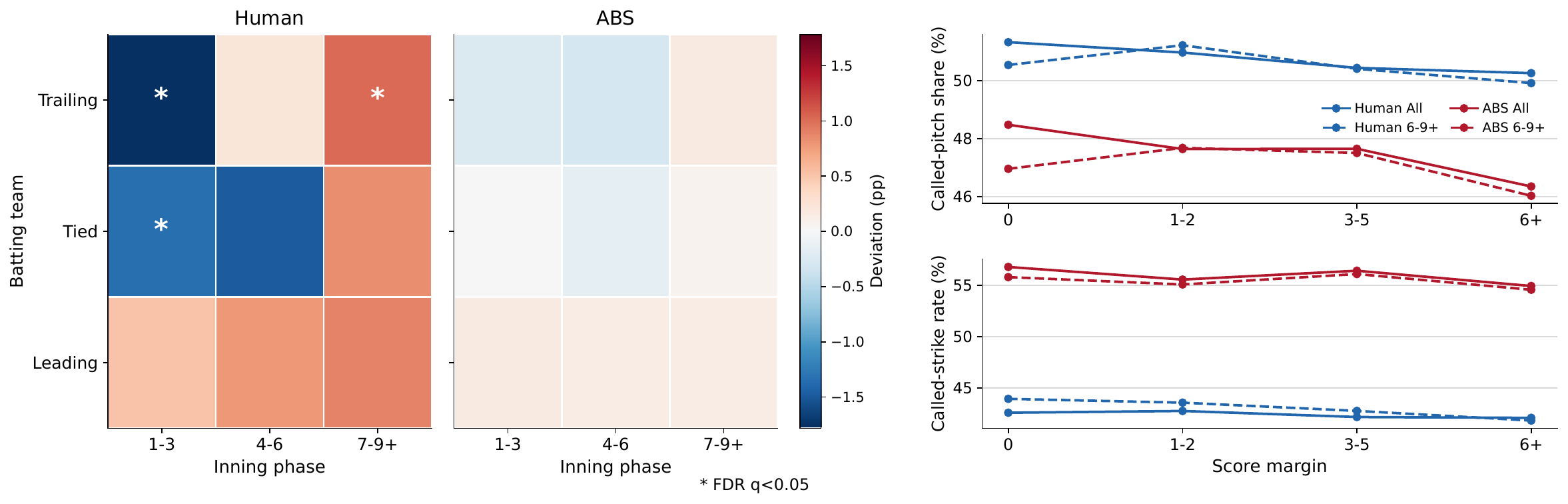}
\caption{\textbf{Game phase, score state, and selection diagnostics.} Left: phase-by-score-state deviations from each period's average called-strike probability after location and count controls, grouped by whether the batting team is trailing, tied, or leading. Right: score-margin diagnostics for called-pitch share and called-strike rate among called boundary-band pitches. Asterisks indicate FDR \(q<0.05\).}
\label{fig:h3_phase_score}
\end{figure}

Figure~\ref{fig:h3_phase_score} explains why the go-home label is too blunt. The human-period heatmap shows that the phase effect depends on score state. Early trailing plate appearances were strongly less strike-prone (-1.94 pp, FDR \(q=0.00049\)), while late trailing plate appearances were more strike-prone (+1.01 pp, FDR \(q=0.025\)). Early tied situations were also less strike-prone (-1.35 pp, FDR \(q=0.0077\)). Leading-team cells were more weakly estimated and did not survive FDR correction. In the ABS seasons, none of the phase-by-score cells survived FDR correction. The pattern is therefore better described as game-phase drift than as a uniform late-game zone expansion.

The selection diagnostics provide an important guardrail. The share of pitches that become called pitches and the called-strike rate among boundary-band called pitches vary with score margin. This means that score-state cells should not be read as pure umpire preferences without caution: batters, pitchers, and catchers may select into different taken-pitch situations depending on the game state. At the same time, the displayed selection diagnostics do not overturn the main result. They caution against overinterpreting individual score-margin cells, but the broader phase pattern remains: in the human period, early innings were less strike-prone and late innings were more strike-prone; in complete ABS seasons, the same structure was not present.

The verdict for Audit 3 is partial support with an important qualification. The data support the general suspicion that the human strike zone changed over game phase, but they do not support a simple go-home rule. Human umpires did not appear to expand the zone uniformly whenever the game was late or lopsided. Instead, the evidence points to a modest late-game drift concentrated in close or pressure-sensitive contexts. This makes Audit 3 weaker than count balancing, but still informative: the human strike zone appears to have moved with game context, while the ABS benchmark remains largely flat.

%% file: sec/6_analysis4.tex
\subsection*{Audit 4: Do catchers and pitchers leave residual strike-zone signatures?}
\label{sec:analysis_identity}

The fourth audit target is the familiar claim that some catchers can ``steal'' strikes. In baseball terms, this claim predicts that otherwise similar borderline pitches should receive different called-strike probabilities depending on the receiver. We also test a related but weaker claim: whether pitcher identity itself leaves a residual called-strike signature after accounting for pitch location and context. If catcher presentation or battery identity affected human calls, residual variation should be visible in the human-umpire period and should attenuate when ball-strike adjudication is automated.

This analysis is not a raw leaderboard of catchers or pitchers. Catchers and pitchers are not randomly assigned to each other, and both are embedded in team, umpire, and game contexts. Individual estimates can also be sensitive to sparse cells. For that reason, we emphasize shrinkage-adjusted residual dispersion and FDR-supported effects, not raw called-strike rates. The models adjust for pitch location, count, pitch type, handedness, season, umpire, and opposite battery or player controls before estimating identity-level residuals.

\begin{table}[!ht]
\centering
\small
\begin{adjustbox}{width=0.6\linewidth,center}
\begin{tabular}{|l|l|r|r|r|r|}
\hline
\textbf{Role} & \textbf{Period} & \textbf{Entities} & \textbf{Eligible} & \textbf{Shrinkage SD (pp)} & \textbf{Heterogeneity \(p\)} \\ \hline
Catcher & Human & 56 & 39 & 1.13 & \(2.24\times10^{-15}\) \\ \hline
Catcher & ABS & 63 & 43 & 0.00 & 0.774 \\ \hline
Pitcher & Human & 360 & 207 & 1.39 & \(1.29\times10^{-14}\) \\ \hline
Pitcher & ABS & 440 & 259 & 0.12 & 0.075 \\ \hline
\end{tabular}
\end{adjustbox}
\caption{\textbf{Identity-level residual heterogeneity.} Eligible entities have at least 100 called pitches in the main 0.25-ft boundary-band sample. Shrinkage SD is the standard deviation of empirical-Bayes residual effects among eligible entities.}
\label{tab:h4_identity_dispersion}
\end{table}

Table~\ref{tab:h4_identity_dispersion} shows the clearest version of the catcher result. In the human period, eligible catchers had a shrinkage-adjusted residual standard deviation of 1.13 percentage points, with strong evidence of heterogeneity. Under ABS, the corresponding catcher dispersion was effectively zero and not statistically supported. This is the pattern the audit would expect if catcher-associated receiving affected human calls but not automated calls. The key result is not that one particular catcher ranked first or last; it is that catcher identity explained residual called-strike variation under human umpires and stopped doing so under ABS.

\begin{figure}[!ht]
\centering
\begin{minipage}[t]{0.4\linewidth}
\centering
\includegraphics[width=\linewidth]{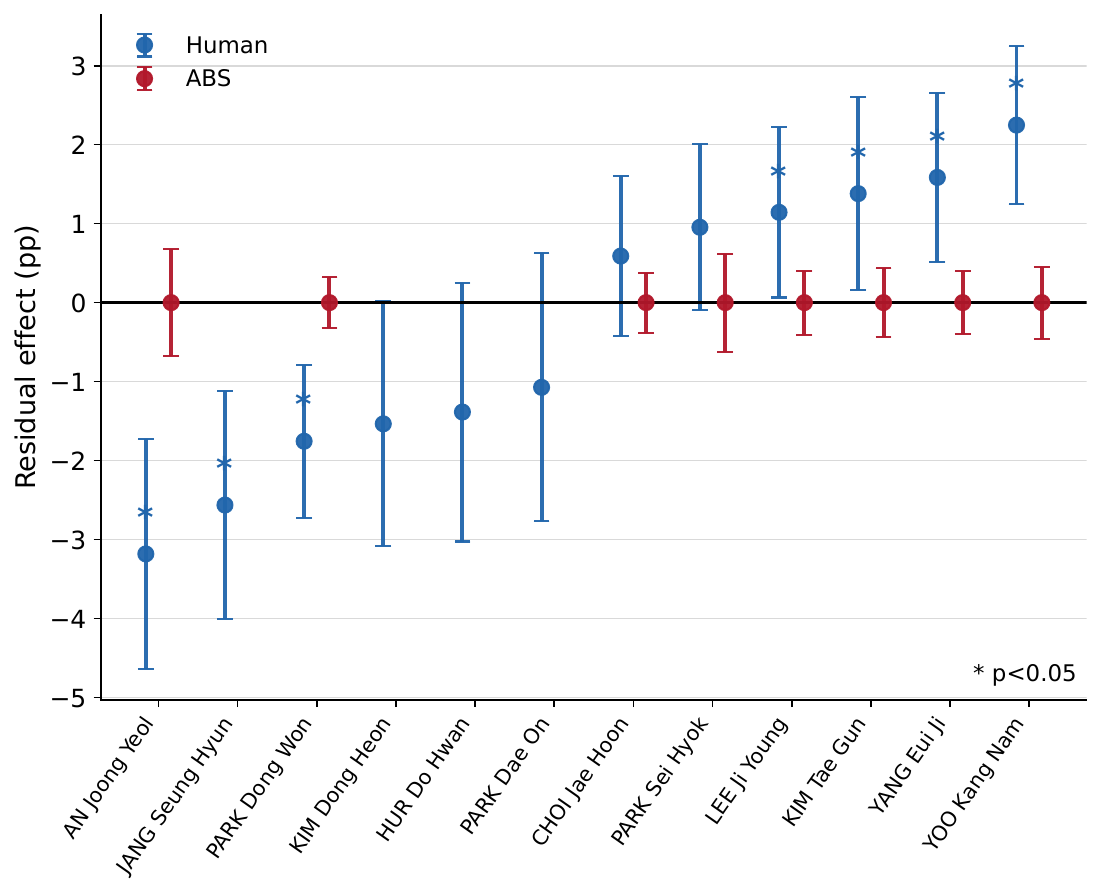}
\textbf{(a) Catchers}\\[0.4em]
\end{minipage}
\begin{minipage}[t]{0.4\linewidth}
\centering
\includegraphics[width=\linewidth]{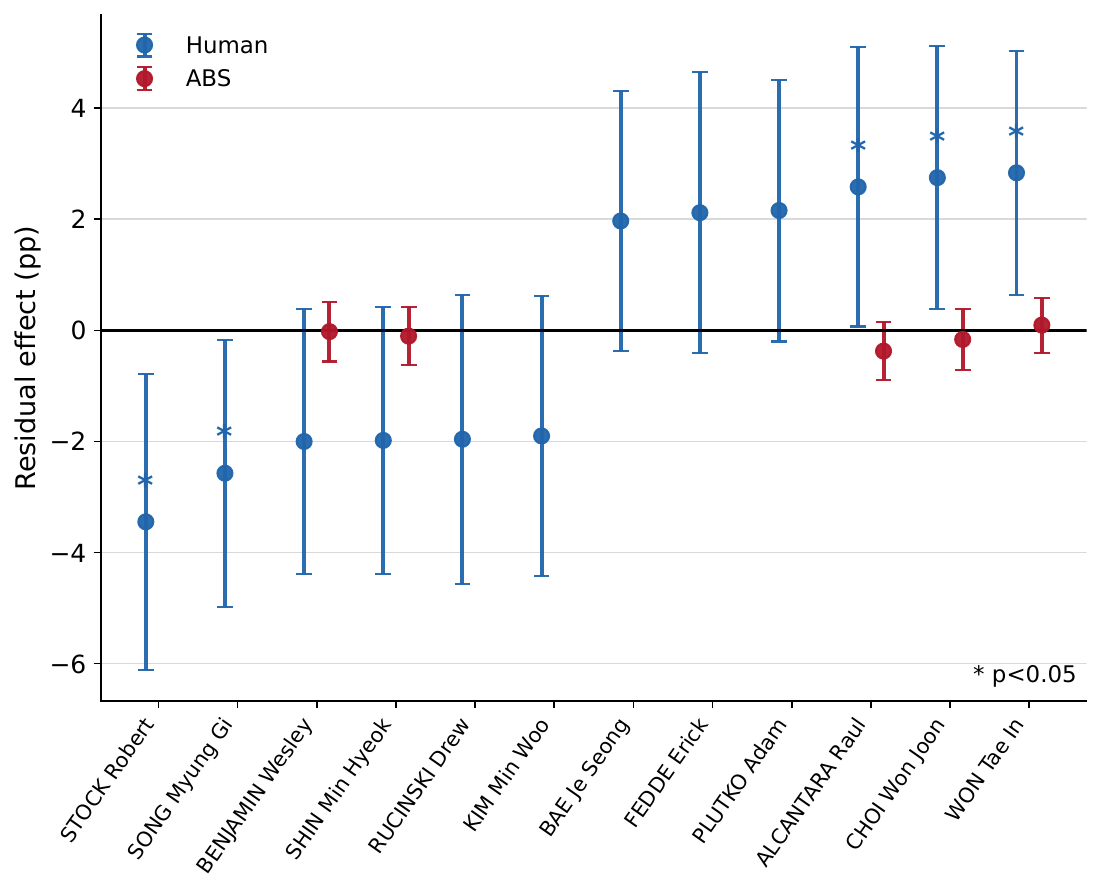}
\textbf{(b) Pitchers}\\[0.4em]
\end{minipage}
\caption{\textbf{Battery-identity residual called-strike effects.} Points show shrinkage-adjusted catcher and pitcher effects for figure-eligible entities before and after ABS. Positive values mean more called strikes than expected after model controls; negative values mean fewer. Asterisks indicate FDR \(q<0.05\). Pitcher cells are sparser than catcher cells, so individual pitcher estimates are interpreted more cautiously.}
\label{fig:h4_identity_effects}
\end{figure}

Figure~\ref{fig:h4_identity_effects} gives the same result at the individual-residual level. Several catcher residuals survive FDR correction in the human period, with figure-eligible catchers spanning roughly five percentage points from the most negative to the most positive residual effects. Under ABS, the same catcher effects collapse toward zero. This makes the catcher result substantively interpretable as a receiving-related human-period trace: comparable borderline pitches were not called identically across catchers before automation, but catcher identity no longer mattered once the call was automated.

The pitcher result is more diffuse. At the aggregate level, human-period pitcher residual dispersion is statistically detectable, and it also attenuates sharply under ABS. However, the individual pitcher estimates are less stable than the catcher estimates. Pitcher cells are more fragmented because each pitcher contributes fewer called borderline pitches than a regular catcher receives, and pitcher identity is more tightly connected to pitch mix, command, catcher pairing, and batter matchups. In the figure-eligible set, individual pitcher effects should therefore be interpreted as exploratory rather than as evidence that particular pitchers reliably received a personal zone.

The verdict for Audit 4 is strong support for a catcher-associated human-period trace and weaker support for pitcher-associated residual heterogeneity. The catcher result aligns with the common claim that receiving can influence human ball-strike calls: catcher residual variation exists before ABS and disappears under automated adjudication. The pitcher result points in the same aggregate direction but does not justify strong individual pitcher claims. Overall, this audit supports the broader conclusion that some identity-linked structure existed in human calls, especially through catcher identity, but that this structure was largely removed by ABS.

%% file: sec/7_analysis5.tex
\subsection*{Audit 5: Do home teams get calls?}
\label{sec:analysis_home}

The final audit target is the familiar home-advantage claim: home teams may receive more favorable ball-strike calls. Because the outcome is a called strike against the batter, a negative home-batter residual contrast means that home batters received fewer called strikes than comparable away batters, and therefore a more favorable zone. We examine this claim in two stages. First, we test whether individual umpires show systematic home/away residual contrasts. Second, because home context may be more localized than a league-wide crowd effect, we conduct an exploratory screen for unusual umpire--team pairings.

\begin{figure}[!ht]
\centering
\includegraphics[width=0.85\linewidth]{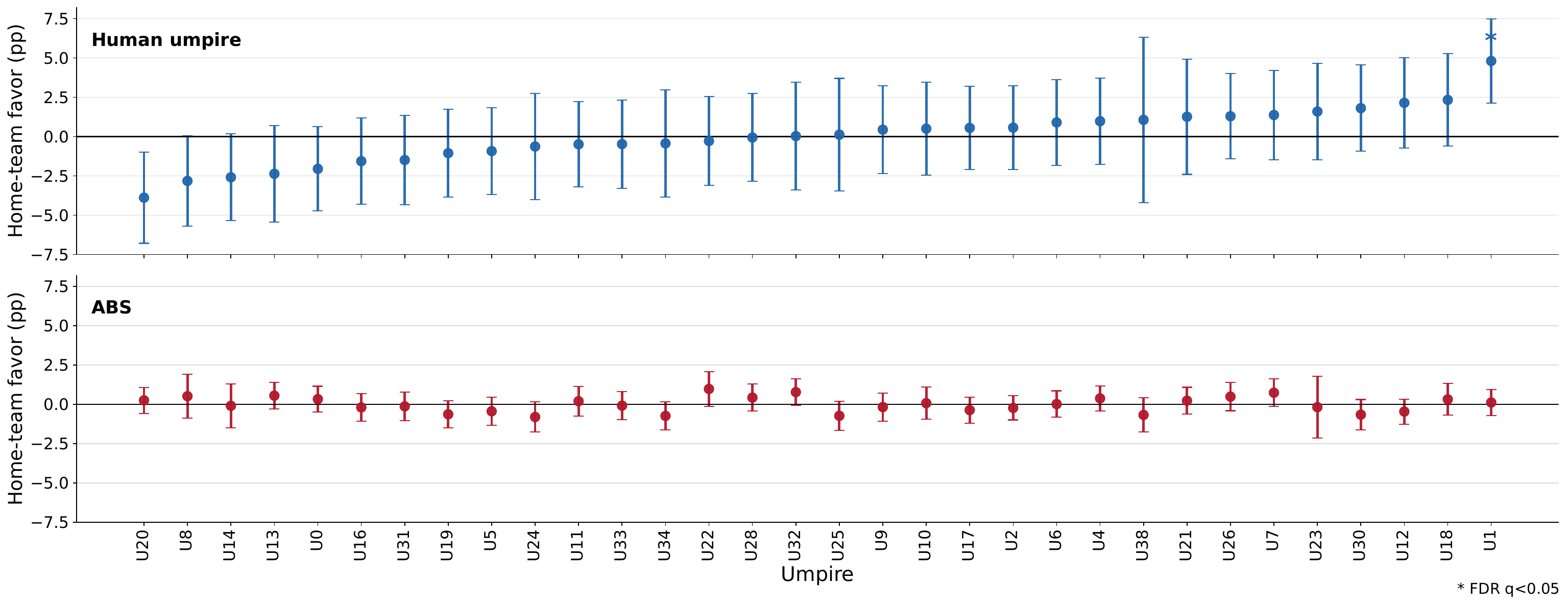}
\caption{\textbf{Umpire-level home/away residual contrasts.} Points show umpire-specific home-batter residual contrasts among umpires observed in both human and ABS periods. Asterisks use FDR-adjusted \(q<0.05\) within period.}
\label{fig:h5_umpire_marginal}
\end{figure}

Figure~\ref{fig:h5_umpire_marginal} shows that the umpire-level home-context test is mostly null. Among 32 eligible umpires observed in both the human and ABS periods, two human-period umpires had nominal \(p<0.05\), but only one survived FDR correction. That umpire was U1, Park Jong-chul. His marginal estimate was -4.80 pp on the called-strike scale, corresponding to a +4.80 pp home-favoring contrast: home batters received fewer called strikes than comparable away batters (\(q=0.014\), \(N=2{,}890\) called pitches across 46 games). No umpire-level home/away contrast was significant under ABS. Thus, the data do not support a broad umpire-level home-zone effect; they identify one human-period exception.

\begin{figure}[!ht]
\centering
\includegraphics[width=0.75\linewidth]{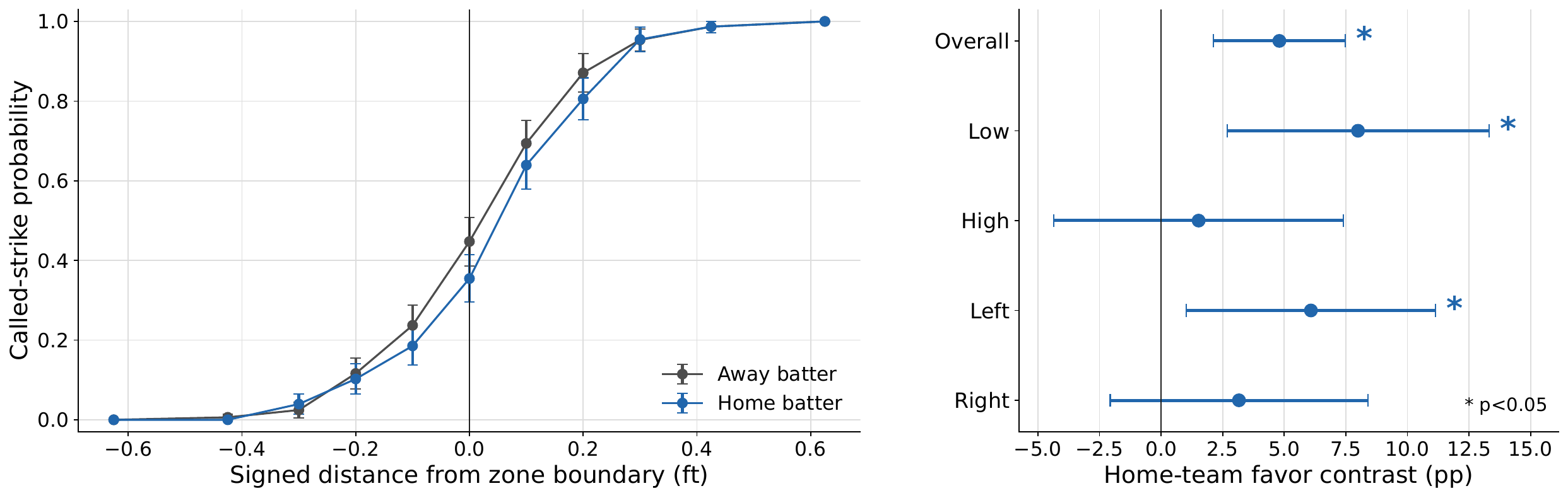}
\caption{\textbf{U1 home/away profile.} The left panel shows raw called-strike probability by signed distance from the rule-zone boundary. The right panel shows adjusted home-favor contrasts by boundary side, where positive values favor home batters. Asterisks indicate \(p<0.05\) in this targeted U1 follow-up.}
\label{fig:h5_u1_profile}
\end{figure}

Figure~\ref{fig:h5_u1_profile} clarifies where the U1 exception comes from. The overall home-favoring contrast is +4.80 pp, but it is not uniform across the strike-zone boundary. It is largest at the lower boundary (+7.99 pp, \(p=0.003\)) and on the left side of the plate (+6.08 pp, \(p=0.018\)), while the high and right-side contrasts are not statistically supported. This pattern is more consistent with a localized boundary asymmetry than with a general home-team shift across the entire zone.

The mostly null umpire-level result does not rule out more localized home-context patterns. A home advantage may not operate as a uniform effect across all umpires and teams; it could appear only in particular umpire--team pairings. We therefore conducted an exploratory screen of eligible umpire-home-team cells. This screen should be interpreted as an audit-lead generator rather than as a confirmatory test. Across 212 eligible cells, 15 were nominally significant, but none survived full-screen FDR correction. Only U1 and U4 produced more than one nominal cell. U1's cells pointed in the same home-favoring direction, consistent with the marginal U1 result above. U4 was different: two of his team-specific cells moved in opposite directions.

\begin{figure}[!ht]
\centering
\begin{minipage}[t]{0.43\linewidth}
\centering
\small
\textbf{(a) Exploratory umpire--team screen}\\[0.5em]
\begin{tabular}{|l|l|}
\hline
\textbf{Screen feature} & \textbf{Result} \\ \hline
Eligible cells & 212 \\ \hline
Nominal leads & 15; none FDR-significant \\ \hline
Repeated nominal umpires & U1 and U4 \\ \hline
U1 pattern & HH +6.64 pp; WO +8.35 pp \\ \hline
U4 pattern & HT -11.36 pp; LT +11.77 pp \\ \hline
\end{tabular}
\end{minipage}
\hfill
\begin{minipage}[t]{0.53\linewidth}
\centering
\textbf{(b) Targeted U4 follow-up}\\[0.5em]
\includegraphics[width=\linewidth]{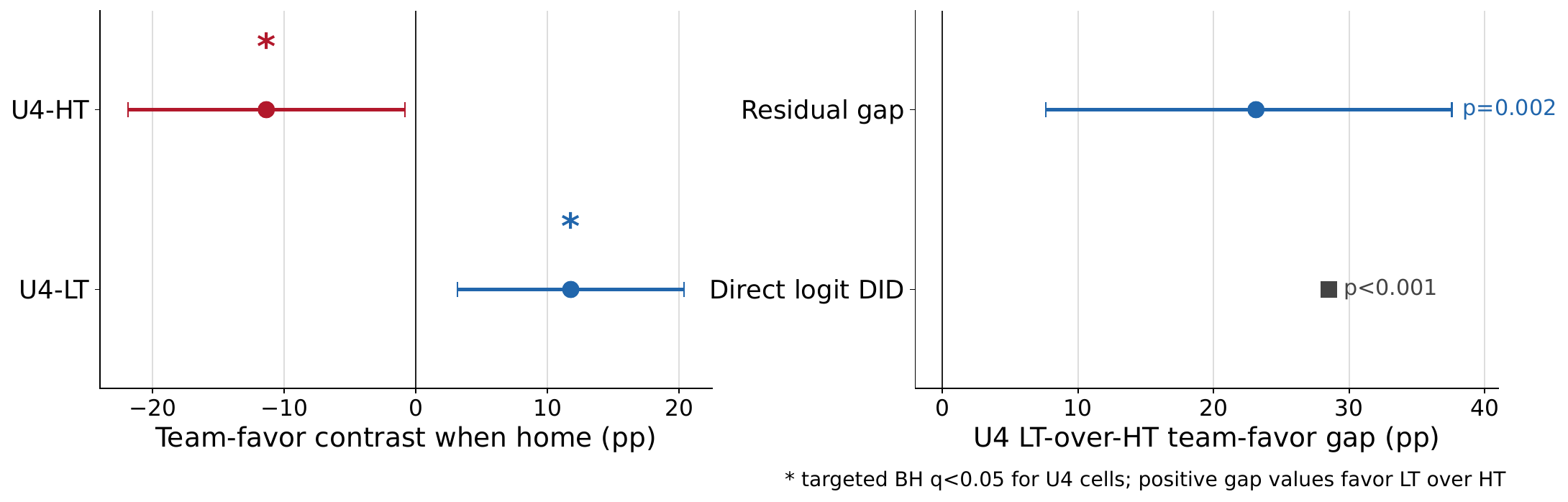}
\end{minipage}
\caption{\textbf{Exploratory umpire--team screen and targeted follow-up.} Panel (a) summarizes the screen that identified repeated nominal umpire--team leads. Positive values mean fewer called strikes for home batters, i.e., a home-favoring contrast. Panel (b) shows the targeted U4 follow-up. The result is not a confirmatory all-cell finding; it is an audit lead for prospective monitoring.}
\label{fig:h5_umpire_team_followup}
\end{figure}

Figure~\ref{fig:h5_umpire_team_followup} summarizes the exploratory screen and the targeted U4 follow-up. Unlike U1, whose repeated nominal cells point in the same home-favoring direction, U4 shows a sharp split between two home-team cells: HT is home-disfavoring, while LT is home-favoring. This opposite-signed pattern is not evidence of a general home advantage, but it is audit-relevant because it suggests a localized umpire--team gap that is much larger than the average home/away pattern.

In the targeted follow-up, the U4 split remains large. On the team-favor scale, the LT-over-HT gap is 23.13 pp with a game-cluster bootstrap interval of 7.61 to 37.57 pp (\(p=0.002\)). A direct logistic interaction gives a similar marginal difference of 28.50 pp (\(p<0.001\); cluster-robust \(p<0.001\)). These estimates are large, but they should be interpreted cautiously. The cells were selected after the exploratory screen and involve 591 pitches across nine games. The result is therefore not a confirmed league-wide team-bias claim. It is better understood as the kind of localized pattern that an automated audit system can surface for prospective monitoring.

The verdict for Audit 5 is mostly null with localized audit leads. The data do not support a general KBO home-zone effect at the umpire level, and no comparable home/away pattern appears under ABS. The U1 result shows that an individual umpire can exhibit a localized home-favoring contrast, especially near particular boundary sides. The U4 follow-up shows that specific umpire--team pairings can produce larger local gaps, but those results are exploratory and should be used to motivate future monitoring rather than to establish a broad home-bias conclusion. Compared with count balancing and catcher-associated residual variation, home-context evidence is weaker, less systematic, and more localized.

%% file: sec/8_relwork.tex
\section*{Related work}
\label{sec:relwork}

\subsection*{Pitch tracking and strike-zone modeling}

Modern ball-strike research builds on pitch-tracking systems such as PITCHf/x, Statcast, and related camera- or sensor-based infrastructure, which made the strike zone a pitch-level measurement problem~\cite{fast2010heck,lage2016statcast,mlbBaseballSavant}. These data have supported analyses of pitch location, movement, release, pitch quality, and zone geometry~\cite{nathan2012determining,healey2017using,whiteside2016ball,hsieh2018graphic}. They also make clear why contextual analyses require location controls: near the zone boundary, small differences in pitch distribution can create large differences in raw called-strike rates. Strike-zone models have moved from simple rectangular comparisons to probabilistic, geometric, and hierarchical approaches~\cite{chen2010contour,hunter2018new,deshpande2017hierarchical,flannagan2024psychophysics}. Our design follows this literature by avoiding raw call comparisons. We focus on boundary-band taken pitches and estimate location-controlled called-strike models with boundary distance, rule-zone status, pitch characteristics, handedness, and season controls.

\subsection*{Umpire accuracy and contextual bias}

Umpire research can be divided into accuracy studies and decision-context studies. Accuracy work asks whether calls agree with a rule-zone or tracking benchmark; contextual-bias work asks whether non-location features affect calls after pitch geometry is considered~\cite{fesselmeyer2021impact,hamrick2015connection,tainsky2015further}. This paper belongs to the latter category. We do not treat every deviation from a modeled zone as an error. Instead, we test whether deviations are structured by count, game phase, player status, catcher or pitcher identity, and home context. Contextual bias need not imply intent. In high-speed perceptual decisions, thresholds may shift with expectations, consequences, or salience. Count pressure is the cleanest example: a marginal strike in a 3--0 count prevents a walk, while a marginal strike in an 0--2 count can end the plate appearance. This asymmetry motivates our count-balancing audit.

\subsection*{Catcher framing, status, and home context}

Catcher framing studies show that receiver identity can be associated with changes in called-strike probability on similar pitches~\cite{deshpande2017hierarchical,leblanc2021impact,judge2018bayesian}. Catcher identity is therefore both a potential confounder and a substantive object of study. Our catcher analysis estimates residual catcher-associated effects after location and context controls, while avoiding claims about direct glove mechanics or intent. Sports-officiating research has also examined status, reputation, crowd pressure, and home advantage~\cite{macmahon2008contextual,tainsky2015further,hamrick2015connection}. Baseball is useful for these questions because of its large pitch-level sample, but difficult because status is hard to measure and home effects may be weaker for highly repeated, closely monitored ball-strike decisions. We therefore treat salary-based status tests as proxy-sensitive and home-context tests as primarily null-oriented.

\subsection*{Automated officiating as audit infrastructure}

Automated ball-strike systems are often discussed as replacements for human umpires or as accuracy-improving tools~\cite{nbcnewsTechnicalDifficulties,espnWhenWill,nytimesTripleAGames,dongaKoreanBaseball}. Prior KBO ABS work has compared aggregate human and automated strike zones. Our use of ABS is different: the automated era serves as a diagnostic benchmark for auditing contextual human judgment. Patterns that are large under human umpires but attenuate under ABS are more consistent with human context-sensitive adjudication than with pitch distribution alone; patterns that persist under ABS are interpreted more cautiously. This does not make ABS a randomized counterfactual. Rosters, strategies, and player behavior change across seasons. Still, automation provides a stronger benchmark than comparing one human-umpire regime with another. The contribution of this study is to use that benchmark to map which suspected contextual effects in human ball-strike calls are strong, weak, or unsupported.

%% file: sec/9_discussion.tex
\section*{Discussion}
\label{sec:discussion}

The main finding is not that human umpires were biased everywhere. It is that some suspected contextual effects were strong, some were modest, and others were weak or mostly absent. Count pressure is the clearest case: before ABS, borderline pitches were more strike-prone in 3--0 counts and much less strike-prone in two-strike counts, while the same count-state structure was nearly absent under ABS. This provides the strongest evidence that human ball-strike adjudication was context-sensitive beyond pitch location alone.

Other audits produce more qualified evidence. Game phase shows a modest human-period drift, with calls becoming less strike-prone early and more strike-prone late, especially in late-close situations. However, the pattern does not support a simple ``go-home'' rule. Status effects are weaker and proxy-sensitive, especially because salary only imperfectly measures reputation. Catcher identity provides a stronger secondary signal: catcher-associated residual variation is present under human umpires and collapses under ABS. Home-context evidence is mostly null, with only localized human-period exceptions.

This hierarchy matters. The same framework that detects large count effects and catcher-associated residual structure does not mechanically find bias everywhere. That strengthens the interpretation of the positive results and clarifies where audit attention is most useful. Count pressure is the most actionable target for umpire training, feedback, or post-game review. Catcher-associated residuals can be monitored as a receiving-related signal without reducing the analysis to individual blame. Home-context results suggest that broad home-zone bias is less central, although localized audit leads may still justify prospective monitoring.

The findings also have implications beyond the KBO's full-ABS setting. MLB has moved toward an ABS challenge system rather than fully automated ball-strike calling~\cite{mlbABSChallenge2026,baseballSavantABSDashboard}. In such hybrid systems, the key question is not only whether automated review corrects missed calls, but how players, catchers, pitchers, and umpires interact around the right to challenge. The biases documented here may shape which pitches players choose to challenge, when they hesitate, and how human umpires behave when their calls can be selectively reviewed. Future work should therefore study contextual bias not only as a property of human calls, but also as part of a human--automation interaction loop.

Overall, automated officiating should be understood not only as replacement infrastructure but also as audit infrastructure. ABS provides a benchmark for identifying where human rule enforcement was most context-sensitive before automation, while also showing where common baseball suspicions receive little support.

\section*{Limitations}

This study is observational. Count, inning, score, player identity, catcher assignment, and home context are not randomly assigned, so location controls, zone controls, count controls, and season fixed effects cannot eliminate all confounding from pitch selection, batter approach, pitcher command, or team strategy. ABS is also a diagnostic benchmark rather than a randomized counterfactual: the automated period differs from the human period in rosters, strategies, league conditions, and player adaptation. The analysis is conditioned on taken pitches, so the sample is downstream of swing decisions that may vary across counts, pitchers, and game states. Key variables are also imperfectly measured: batter-specific zone fields are not direct observations of perceived umpire zones or exact operational ABS boundaries, salary is only a proxy for reputation, and catcher identity is reconstructed from lineup data rather than pitch-by-pitch defensive tracking. These limits, together with multiple contextual tests, are why we distinguish strong, secondary, exploratory, and null findings and why replication in additional seasons or leagues is needed.

%% file: sec/10_conclusion.tex
\section*{Conclusion}
\label{sec:conclusion}

The KBO's transition to ABS provides a benchmark for auditing long-standing claims about contextual human umpiring. The strongest evidence is count pressure: before ABS, borderline pitches were more strike-prone in three-ball counts and much less strike-prone in two-strike counts, while the same count-state structure was nearly absent under ABS. Smaller but meaningful human-period patterns appear in late-game contexts and catcher-associated residual variation. By contrast, salary/status evidence is proxy-sensitive, and average home-team advantage is mostly unsupported. This hierarchy is the paper's main empirical message. Human ball-strike calls were not context-sensitive everywhere, but they were context-sensitive in structured and baseball-interpretable ways. Automated officiating should therefore be understood not only as a replacement for human judgment, but also as audit infrastructure: a benchmark for identifying where human rule enforcement was most vulnerable to contextual drift and where common suspicions receive little support.

\section*{Data availability}

The pitch-level data used in this study were collected from publicly accessible Naver Sports game records~\cite{naver}. The manuscript is based on derived analysis tables, scripts, and figures in the project repository. Cleaned derived tables and replication scripts will be released subject to data-source terms of use.